\documentclass[aps,amsmath,amssymb,twocolumn,showpacs,superscriptaddress]{revtex4}%
\usepackage{amsmath}
\usepackage{xcolor}
\usepackage{graphicx}% Include figure files
\usepackage{dcolumn}% Align table columns on decimal point
\usepackage{bm}% bold math
\usepackage{float}
\usepackage{epsfig}% FOR CORRECT FIGURE CAPTIONS
\usepackage{graphicx}%
\usepackage{amsmath}%
\usepackage{amsfonts}%
\usepackage{amssymb}
\usepackage{natbib}
\usepackage{soul}
\usepackage{mhchem}
\usepackage{color}
\DeclareMathOperator{\sech}{sech}
\DeclareMathOperator{\sinhc}{sinhc}
\DeclareMathOperator{\sgn}{sgn}
\newcommand{\ora}{\overrightarrow}
\newcommand{\bE}{{\bf E}}
\newcommand{\Ups}{{\bf\Upsilon}_m^{(j)}(r,z,\phi)}
\newcommand{\Upsp}{{\bf\Upsilon}_{m'}^{(j')}(r,z,\phi)}
\usepackage{amssymb}
\usepackage{scalerel}
\usepackage{appendix}
\usepackage{graphicx}
\begin{document}
\title{Reversible Self-Replication of Spatio-Temporal Kerr Cavity Patterns}
%\title{Reversible Kerr Comb Self-Replications in Cylindrical Microresonators}
% Frequency Comb Replicas in Cylindrical Micro Resonators
\author{Salim B. Ivars}
\affiliation{Institut Universitari de Matem\`{a}tica Pura i Aplicada, Universitat Polit\`{e}cnica de Val\`{e}ncia, 46022 Val\`{e}ncia, Spain}
\affiliation{ICFO--Institut de Ciències Fotòniques, The Barcelona Institute of Science and Technology, 08860 Castelldefels (Barcelona), Spain}
\affiliation{Departament de F\'{i}sica, Universitat Polit\`{e}cnica de Catalunya, 08222 Terrassa (Barcelona), Spain}
\author{Yaroslav V. Kartashov}
\affiliation{ICFO--Institut de Ciències Fotòniques, The Barcelona Institute of Science and Technology, 08860 Castelldefels (Barcelona), Spain}
\affiliation{Institute of Spectroscopy, Russian Academy of Sciences, Troitsk, Moscow, 108840, Russia}
\author{Lluis Torner}
\affiliation{ICFO--Institut de Ciències Fotòniques, The Barcelona Institute of Science and Technology, 08860 Castelldefels (Barcelona), Spain}
\affiliation{Universitat Polit\`{e}cnica de Catalunya, 08034 Barcelona, Spain}
\author{J. Alberto Conejero}
\affiliation{Institut Universitari de Matem\`{a}tica Pura i Aplicada, Universitat Polit\`{e}cnica de Val\`{e}ncia, 46022 Val\`{e}ncia, Spain}
\author{Carles Mili\'{a}n}
\email{carmien@upvnet.upv.es}
\affiliation{Institut Universitari de Matem\`{a}tica Pura i Aplicada, Universitat Polit\`{e}cnica de Val\`{e}ncia, 46022 Val\`{e}ncia, Spain}
\affiliation{ICFO--Institut de Ciències Fotòniques, The Barcelona Institute of Science and Technology, 08860 Castelldefels (Barcelona), Spain}
%
%%%%%%%%%%%%%%%%
% NEW ABSTRACT %
%%%%%%%%%%%%%%%%
%
\begin{abstract}
We uncover a novel and robust phenomenon that causes the gradual self-replication of spatiotemporal Kerr cavity patterns in cylindrical microresonators. These patterns are inherently synchronised multi-frequency combs. Under proper conditions, the axially-localized nature of the patterns leads to a fundamental drift instability that induces transitions amongst patterns with a different number of rows. Self-replications, thus, result in the stepwise addition or removal of individual combs along the cylinder's axis. Transitions occur in a fully reversible and, consequently, deterministic way. The phenomenon puts forward a novel paradigm for Kerr frequency comb formation and reveals important insights into the physics of multi-dimensional nonlinear patterns.
\end{abstract}
%%%%%%%%%%%%
% OLD ABSTRACT %
%%%%%%%%%%%%
%
%\begin{abstract}
%We uncover a novel and robust phenomenon that causes the gradual self-replication of spatiotemporal patterns in cylindrical Kerr microresonators. The axially localized nature of these pattern excites a fundamental drift instability landscape that induces transitions amongst hexagonal patterns with a different number of spots. The transitions occur in a full reversible and, consequently, deterministic way. The phenomenon reveals important insights about the physics of multi-dimensional nonlinear patterns and may be relevant for applications requiring multiple synchronised frequency comb sources.
%\end{abstract}
\maketitle
%
%%%%%%%%%%%%%%%%%%%%%
% NEW- INTRODUCTION %
%%%%%%%%%%%%%%%%%%%%%
%\textit{Introduction.---}

% MESSAGE Kerr cavity solitons --> Boom in combs, which are of very broad interest, as they directly impact a wide range of applications such as metrology, sensing, ranging, astronomy, and communications, amongst others [REFS]. 

%Other 1D related nonlinear waves such as Turing rolls and variations of it --> additional versatility into comb related foundamental phenomena and tecnological interest. Frequency comb formation research remains, though, restricted to essentially one-dimensional geometries, i.e., single or few moded microrings.

The demonstration of a microresonator temporal soliton \cite{herr14} and its subsequent stabilization \cite{yi16,brasch16} \iffalse {,guo17} \fi yielded and strongly boosted a wide range of applications of the associated stable frequency combs, such as frequency synthesis \cite{spen18}, spectroscopy \cite{suh16}, communications \cite{marinNAT}, and ranging \cite{suh18}, amongst many others (see, e.g., Refs. \cite{gaetRev,gaetSC,pasqPR} for reviews). Aside from the widely employed single soliton states, the interest in other nonlinear waves such as Turing rolls is also growing rapidly as these waves are also very useful and may result in more efficient frequency comb generation \cite{coleNP,karpov19,pasq20,szab20,chembo14,parraPRE18,menyuk20}. Nevertheless, ongoing research on microresonator frequency combs remains strongly focused on essentially one-dimensional geometries, while potential advantages or qualitatively new ways to control and manipulate combs in multi-dimensional cavity geometries are not yet clearly identified.

% In this Letter, we show that frequency combs can be dterministically and stepwise replicated or erased along the z-axis of microcylinders. yieldiong multi-frequ7ency comb states, cooresponding in our particular case to localized Hexagonal Kerr-cavity patterns.
% this effect motivates a novel paradigm for frequency comb formation. In addition, the effect presented is based on a very rare type of pattern transoformation, very little explored til date. 

%present a novel and intrinsically two-dimensional effect in microcylinders, representing the latter the natural two-dimensional extension of microrings. We
%%%%%%%%%%%%%
%%% FIG.1 %%%
%%%%%%%%%%%%%
\begin{figure*}
\begin{center}
\includegraphics[width=.92\textwidth]{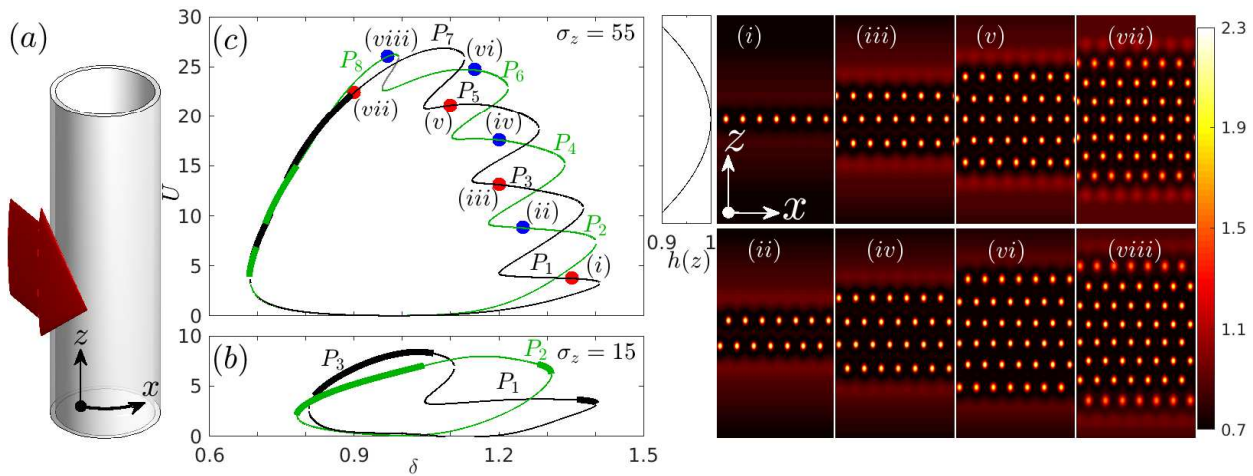}
\caption{(a) Sketch of the driven micro-cylinder. (b), (c) snaking diagrams of pattern families $P_{odd}$ and $P_{even}$ on the $(\delta,U)$ plane for $\sigma_z=15$, $\sigma_z=55$, respectively. Thick traces denote stability. Dots in (c) correspond to the profiles, $|\psi(x,z)|$, shown in insets $(i)-(viii)$. Panels' areas span over $x\in[-16,16)$ (full circumference, $2\pi R$), $z\in[-20, 20]$. Here, $h_0=\xi(x)=1$. Scaling: $\Delta\delta=0.1\Leftrightarrow15.6$ MHz, $\sigma_z=10\Leftrightarrow4.9$ mm, $\Delta z=40\Leftrightarrow 1.96$ cm, $h=1\Leftrightarrow$46 mW/mm. \label{f1}}
\end{center}
\end{figure*}
In this Letter, we show that cylindrical Kerr microresonators, the natural two-dimensional extension of microrings [see Fig.\ref{f1}(a)], offer a remarkably robust mechanism to replicate and erase frequency combs along the axial direction. Each comb corresponds to a row of a spatiotemporal hexagonal pattern constituted by a periodic arrangement of solitonic pulses, so that the $N$-row spatiotemporal patterns localized along the axial direction are regarded as $N$-frequency comb states [c.f. Fig.\ref{f1}]. Individual combs are added or removed one by one in a step-wise deterministic way solely by tuning the width of the external laser pump beam. A powerful feature of the multi-comb states introduced here is that all combs tend to be identical to each other, thus matching among their free spectral range (FSR) occurs naturally, a property that is of great importance in the areas of high precision dual-comb spectroscopy \cite{hansch}, ultra-fast communications \cite{marinNAT}, and radio-frequency (RF) links \cite{pavlovOL}. In addition, such combs are inherently synchronised, which is fundamentally attractive \cite{gaetSync} and may be beneficial for imaging applications requiring many combs \cite{bao19}. \iffalse ,{erkSync} \fi%Applications requiring $N$-combs may benefit from hostiing them into a single monolithic cavity, which may lead to smaller footprints of $N$-comb sources. In addition, not taht cylinders's driven by variable pump width behave, effectively, as cavities of on demand height, meaning that cylinders may be opearted as standard microrings or as highly multimodal cavities.
In practice, multi-frequency comb states may be realised with microrods \cite{delhaye}, pumped from rectangular waveguide or flattened fibers, or integrated microcylinders \cite{kipp06} coupled to integrated waveguides.
%In addition to the above, note that real time reconfigurability of multi-comb states fully avoids intermediate chaotic regimes that may severely compromise the combs' lifetimes.

From a fundamental standpoint, the effect uncovered here is related to the phenomenon of pattern self-replication, which typically manifests as spot multiplications in reaction-diffusion systems \cite{selfrepNAT,pears,reyn94} or pattern expansions in thermoconvection \cite{boden91}. Self-replications include the symmetry-preserving transformations occurring within a given pattern family, as is the case here, in contrast to symmetry transformations, widely studied in optics and other contexts \iffalse, ranging from optics \cite{martin96,ackeman95} to biology \cite{mullin07}  (see, e.g., Refs. \fi \cite{mikhPR06,residori05,arecchiPR}. \iffalse for reviews)\fi %pisarch14}. 
Replication phenomena are till date regarded as uncontrollable expansions %driven by unstable wall modes 
\cite{knob_SH_review} the taming of which represents a fundamental cornerstone yet to be achieved. In sharp contrast with previously known mechanisms \cite{boden91,knob_SH_review}, we show that the spatiotemporal patterns emerging after addition or removal of the entire new rows remain locked. The key ingredients for such transformations are the existence of drift instability %(see, e.g., Refs. \cite{scrCS1,scrCS2,scr05apb,parraPRL}) % scrCS2, parraPRL --> out
and intertwined families of non-linear waves %, e.g., Refs.\cite{barash99,parraPRE18,knob10,knob_SH_review}) % barash99, 
-- rather general features of dissipative systems. These two features simultaneously present in our system enable stepwise self-replications and self-erasures of the multi frequency comb states. Also, our findings are important for the fundamental understanding of pattern transformations and the physics of boundary effects such as stabilization \cite{kozPRL} and geometrical frustration \cite{residori05}.

%Pattern formation has remained of great fundamental importance since the early days of nonlinear science \cite{crossRMP} for they exhibit universal behaviours associated to their symmetries \cite{hoyle}, represent spontaneous emergence of order far from equilibrium \cite{per}, and often rule the very existence of dissipative solitary waves \cite{woodsPD99,coulletPRL00}.

\begin{figure}
\begin{center}
\includegraphics[width=.49\textwidth]{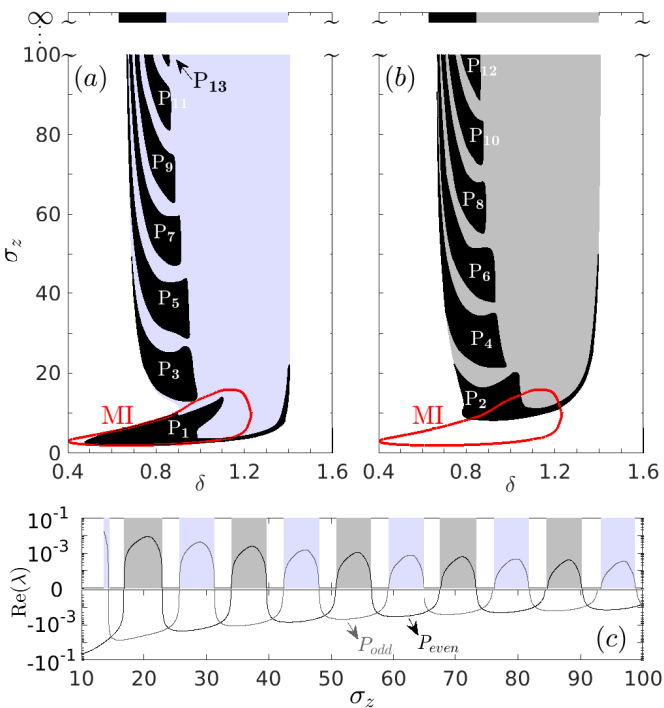}
\caption{Existence and stability charts for (a) $P_{odd}$, (b) $P_{even}$ on the $(\delta,\sigma_z)$ plane [$h_0=1$, $\xi(x)=1$]. Patterns are stable (unstable) within the black (coloured) areas. Domains for extended hexagonal pattern ($\sigma_z\rightarrow\infty$) are also shown. Red contour encloses MI area. (c) Drift eigenvalue vs $\sigma_z$ at $\delta=0.84$ for the patterns in (a) and (b). Areas in (c) [colour matched with (a) and (b)] mark the drift bands for $P_{odd}$ (purple), $P_{even}$ (gray). Scaling: $\Delta\delta=0.1\Leftrightarrow15.6$ MHz, $\sigma_z=10\Leftrightarrow4.9$ mm.\label{f2}}
\end{center}
\end{figure}

\begin{figure*}
\begin{center}
\includegraphics[width=.99\textwidth]{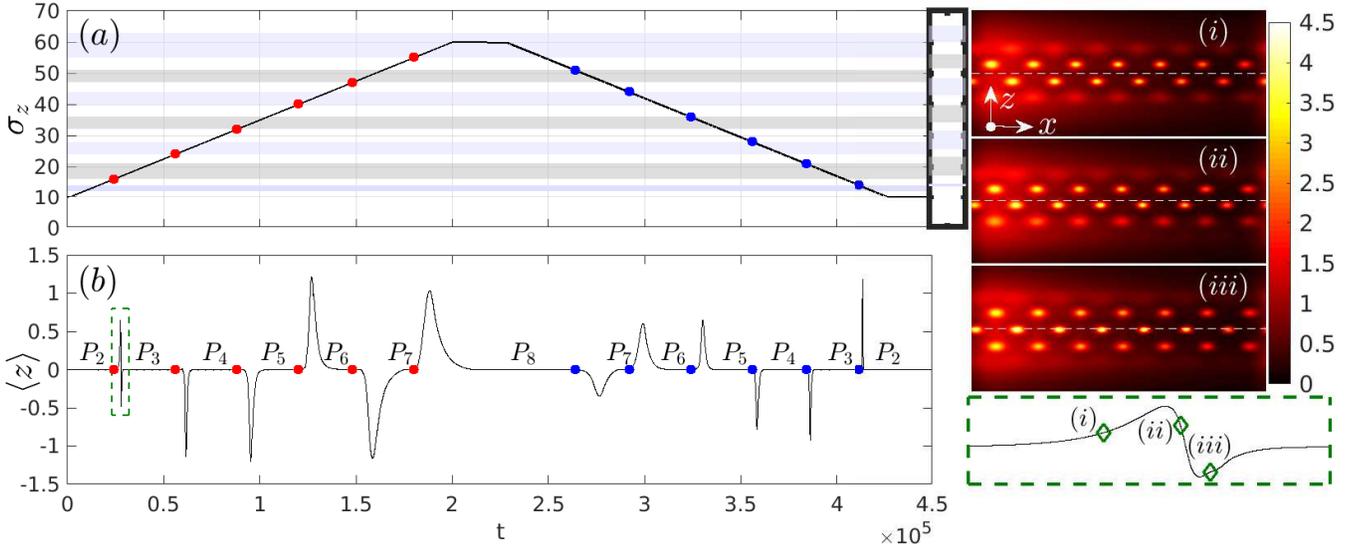}
\caption{Pattern self-replication and self-erasure under the two-dimensional localisation of the pump, $h_0\approx9$, $\sigma_x=2$, and varying $\sigma_z(t)$ with $\delta=0.84$. (a) and (b): evolution of $\sigma_z$ and center of mass. Dots in (a,b) mark the entrance into the drift-instability regions for $P_{odd}$ (purple), $P_{even}$ (gray). Inset in (a) shows the drift bands for $\xi=1$ [c.f. Fig.\ref{f2}(c)]. The dashed green rectangle zooms the $P_2\rightarrow P_3$ transition from (b), while distributions $|\psi(x,z)|^2$ correspond to labels $(i)-(iii)$. Axes of panels $(i)-(iii)$: $x\in[-16,16)$ (full circumference), $z\in[-12, 12]$.
Scaling: $\sigma_z=10\Leftrightarrow4.9$ mm, $\langle z\rangle=1\Leftrightarrow490\ \mu$m, $t=1\Leftrightarrow10^3$ roundtrips, $\tau\approx6.4$ ps, total propagated time is $\sim 2.9$ ms, $\Delta z=24\Leftrightarrow 1.2$ cm, $\sigma_x=2\Leftrightarrow\pi R/8$, $h_0=9\Leftrightarrow 3$ W/mm.\label{f3}}
\end{center}
\end{figure*}

\begin{figure}
\begin{center}
\includegraphics[width=.49\textwidth]{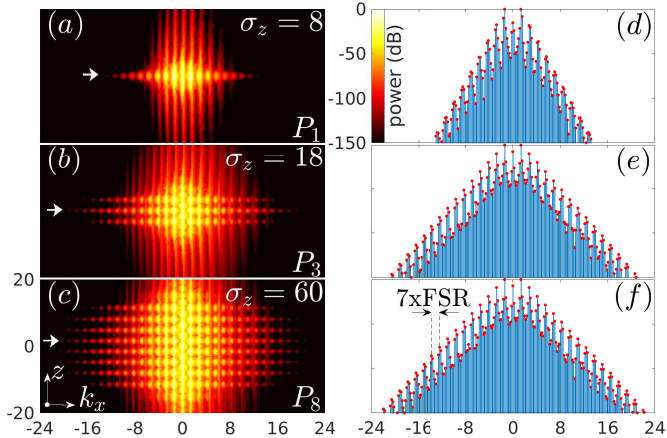}
\caption{Robust multi-comb spectra for (a) $P_1$, (b) $P_3$, (c) $P_8$ ($\sigma_z$ in labels) for the same parameters as Fig.\ref{f3}. (d,e,f), spectra at the selected $z$-positions (white arrows) in (a,b,c), respectively. All power levels are in the range $[-150,0]$ dB. The background field, $\psi_0(x,z)$, was subtracted prior to Fourier transforms.
Scaling: $\Delta\kappa_x=1\Leftrightarrow 1/40$ $\mu \text{m}^{-1}$. At $\lambda_p=1.55$  $\mu$m, $\Delta\kappa_x=48\Leftrightarrow\lambda_0\in[1.35,1.82]\mu$m. $\Delta z=40\Leftrightarrow1.96$ cm.
\label{f4}}
\end{center}
\end{figure}

%Because pattern formation yields organization of large complex systems, their deterministic transformations are paramount phenomena in many areas of science \cite{mikhPR06,residori05,arecchiPR}. %,pisarch14}.
%By far, the most studied transformations concern the selective transitions amongst families with different symmetries, observed in several disciplines such as optics \cite{martin96,ackeman95}, biology \cite{mullin07}, and {thermoconvection} \cite{boden91}. A related, but very different in nature, type of pattern transformations are self-replications In the latter case, self-replications are 
%%%%%%%%%
% MODEL %
%%%%%%%%%
%\textit{Two-dimensional continuous model.---}
%
Applying the modal expansion approach \cite{chembo10,chembo13} to a cylindrical microresonator, in which light orbits around its cross-section and diffracts along its axis [c.f. Fig.\ref{f1}(a)], yields the generalised damped-driven nonlinear Schr\"{o}dinger equation \cite{lugiato87,haeltOC92}:
\begin{eqnarray}
&&\nonumber \partial_t\psi=%-i{\mathcal{D}}_x\psi
\frac{i}{2}\left[\partial_x^2+\partial_z^2\right]\psi-[1+i\delta]\psi+i|\psi|^2\psi+ih(z,x,t),
\\ &&
%{\mathcal{D}}_x\equiv\sum_{m\geq2}\frac{q_{m,x}}{[2b_{2,x}]^{\frac{m}{2}}}(-i\partial_{x})^m,\ \
h=h_0e^{(-z^2/\sigma_z^2)}\xi(x,t),
\label{eq1}
\end{eqnarray}
accounting, respectively, for dispersion, diffraction, losses (unity), cavity-laser detuning, Kerr nonlinearity, and pump (see Supplemental Material for model scaling and derivation \cite{SM}). Similar models can be used to study comb formation in micro-bottles \cite{d2}. Here, we are {primarily} concerned with the effects arising in the multi-frequency comb states due to a variable pump localisation along $z$ [c.f., Figs. \ref{f1}-\ref{f2}].
However, realistic driving beams will typically {couple to} a relatively small region of the cylinder's circumference [c.f. Fig.\ref{f1}(a)], {which is at rest in the lab frame}. Thus the intracavity field, $\psi$, {describing circulating state} and {the localised pump}, $h$, have a huge velocity mismatch $\sim c$. In order to take into account the dynamical effects introduced by their relative motion, which could {have} potentially degraded the practical usefulness of our results [see discussion around Figs. \ref{f3}-\ref{f4}], we will also account for pump localisation in $x$ through the function $\xi(x,t)\equiv\sum_{m=-\infty}^\infty\exp{(-[x-m\Delta x-v_gt]^2/\sigma_x^2)}$, where $\Delta x$ is the normalised circumference,  $x\in[-\Delta x/2,+\Delta x/2)$, and $v_g$ is the normalised group velocity at the pump frequency \cite{SM}. Below, we present our results in normalised units, but we provide the link to a reference geometry, consisting of a hollow silica glass cylinder of $R=200\ \mu$m radius, wall thickness $\approx1\ \mu$m, and quality factor $Q\approx7.6\times10^6$, pumped at $\lambda_p=1.55\ \mu$m. For the sake of estimates only, we considered the pump as a cw state at $\lambda_p=1.55\ \mu$m propagating through a rectangular bus waveguide with a gap of $300$ nm with the cylinder. The pump values will be translated into power per mm along the $z$ direction \cite{SM}.

The patterns we address exist with anomalous dispersion along $x$ and $z$%, $B_{2,x},B_{2,z}>0$
, easily attainable with micro-cylinders. Dispersion along $x$ is readily controlled via the pump's frequency and wall width, and dispersion along $z$ is already anomalous %\cite{YulinOL,laags1,laags2,suchkov17,oreshnOE17,d2,Gorjosab,milianPRL18}
unless modal interactions are specifically engineered \cite{joannoPRL}.

%Here, we consider only the one family of radial modes. The possibility to selectively control the number of such modes via the cylinder's wall thickness, and the unlikely occurrence of (anti-) crossing amongst such families \cite{note_modes}, make this assumption natural. In addition, the cavity modes under consideration are those around the cut-off, i.e., modes with (nearly) zero group velocity along $z$. For these modes, dispersion along $z$ is strongly dominated by GVD, which is in turn typically anomalous \cite{YulinOL,laags1,laags2,suchkov17,oreshnOE17,d2,Gorjosab,milianPRL18}, unless complicated geometries are considered (see, e.g., \cite{joannoPRL}). We note that the continuous GVD term $\sim\partial_z^2\psi$ is obtained in the modal expansion approach \cite{chembo10,chembo13}, strictly, due to the translational invariance of the cavity along $z$.
%
%
%%%%%%%%%%%%%%%%%%%%%%%%%%%%%%%%%
% Patterns and Snaking % % FIG. 1
%%%%%%%%%%%%%%%%%%%%%%%%%%%%%%%%%

%\textit{Patterns and tilted snaking.---}
%
Amongst all possible pattern solutions of Eq. (\ref{eq1}), we focus on hexagonal patterns \cite{firth92} due to their dominant relative stability \cite{tlidi94,tlidi96}. Hereafter, $P_N$ denotes hexagonal patterns with $N$ rows along $z$ and fixed separation $x_p=32/7$ between spots along $x$  [c.f. Figs.\ref{f1}(i)-(viii)]. Stationary patterns, $P_N$, and their stability are computed by imposing $\partial_t\psi=0$ and assuming uniform in $x$ pump ($\xi(x,t)=1$). Figures \ref{f1}(b),(c) show the existence and stability branches as norm $U\equiv\int_0
^{x_p}dx\int_{-\infty}^{\infty}dz|\psi(x,z)-\psi_0(x,z)|^2$ versus cavity detuning for patterns with odd (even) $N$, $P_{odd}$ ($P_{even}$), for $\sigma_z=15$, $\sigma_z=55$ ($\psi_0$ is the background field). % that solves Eq.\ref{eq1} for $B_{1,x}=0$.
Patterns $P_1$ to $P_8$ are shown in insets $(i)$ to $(viii)$. A salient feature of the $U(\delta)$ branches is the \textit{tilted} snaking structure: patterns with larger $N$ are stable and exist at lower $\delta$ values, while patterns with low $N$ exist at higher $\delta$ values, where instabilities typically dominate in $2D$ \cite{firth02,gomila07D}. The gradual shift in $\delta$ of the existence regions is a consequence of the non-uniformity of the pump, $h$, along $z$. Indeed, for uniform in $z$ pump all saddle node bifurcations, i.e., the points where $\partial_\delta U\rightarrow\infty$, are (almost) aligned in $\delta$ yielding \textit{straight snaking} \cite{parraLS,knob10}. While snaking is straightforwardly expected by simple inspection of the pattern profiles \cite{knob10}, the \textit{tilted snaking} is a rare feature (see, e.g., Ref.\cite{firthtilted}) of central importance for this Letter, as it avoids multi-stability within a pattern family and thus enables the pattern self-replicating (-erasure) phenomenon we address below (c.f. \cite{SM} section V).
%
%%%%%%%%%%
% FIG. 2 %
%%%%%%%%%%
%
%%%%%%%%%%
% FIG. 3 %
%%%%%%%%%%
%
%%%%%%%%%%
% FIG. 4 %
%%%%%%%%%%
%
%%%%%%%%%%%%%%%%%%%%
% STABILITY CHARTS % %% FIG. 2- discussion %%
%%%%%%%%%%%%%%%%%%%%

%\textit{Global stability.---}
%
Stability properties of patterns are crucial to elucidate the replication (erasure) process. Thus, all stable domains found in the $U(\delta)$ snaking diagrams [c.f. Figs.\ref{f1}(b),(c)] for $\sigma_z\leq100$ are presented in Figs.\ref{f2}(a), (b) for $P_{odd}$, $P_{even}$, respectively. Stable patterns exist only in the black areas while unstable patterns exist within the light coloured areas: purple (gray) for $P_{odd}$ ($P_{even}$). A crucial feature is that in the interval around $\delta\in[0.75,0.9]$, the \textit{instability} bands of the two families do not overlap, and hence, at a given $\sigma_z$ there exists at least one stable $P_N$. Additionally, unstable bands are characterized by the sole presence of axial drift [exponential] instability, which induces translation of the pattern along $z$ {(upwards or downwards depending on the particular noise seed)}. Note that many types of pattern instabilities exist which could heavily distort and potentially destroy them \cite{gomila07b}. However, regarding Fig. \ref{f2}, they are found only for $\delta\gtrsim0.9$, i.e., outside the interval we address (see \cite{SM} for an example of oscillatory instability).%\textcolor{red}{remove next lines and mention here other dynmical regimes induced by Hopf, etc are found in SM \cite{SM}?(Ref.C Q.2)} For our range of parameters, we did not find any signatures of the wall mode instabilities yielding previously reported uncontrolled expansions \cite{knob_SH_review}.

The axial drift instability, crucial for comb replication, arises in our case due to the non-uniform pump field along $z$. %, which breaks the axial translational symmetry (see, e.g., discussions in  Refs.\cite{sivan08,scr05apb}).
In the uniform pump case, nonlinear states have \textit{neutral} (or Goldstone) internal modes \cite{skryjosab02} associated to displacements along $z$, universally characterised by zero growth rate. %eigenvalue $\lambda$ in the spectrum of the Jacobian of corresponding linear eigenvalue problem and a profile $\propto\partial_z\psi(x,z)$.
However, when the pump is modulated in $z$, the axial translational invariance is broken and the neutral's mode eigenvalue deviates from zero to acquire a positive (negative) real part, thus leading to patterns that are unstable (stable) to drift along $z$ (see, e.g., discussions in  Refs.\cite{sivan08,scr05apb}). In our case, and consistent with formal theorems for conservative systems \cite{sivan08}\iffalse and other works \cite{scrCS1} \fi, the \textit{perturbed} neutral eigenvalue remains real, but oscillates around zero with $\sigma_z$, as shown in Fig.\ref{f2}(c). The regions where $\text{Re}(\lambda)>0$ correspond to drift instabilities for $P_{odd}$ (purple) and $P_{even}$ (gray) pattern families. The amplitude of oscillations of $\text{Re}(\lambda)$ [plotted in $\log$ scale] decreases very fast as $\sigma_z$ increases, because the system tends to recover its axial translational invariance and drift-free dynamics. In the flat pump limit, multi-comb states become infinitely extended hexagonal patterns, whose existence and stability domain is shown in Figs. \ref{f2}(a),(b).% (see Ref.\cite{gomila07b} for further details).

In addition to the above stability properties, we highlight the presence of a modulation instability (MI) region, encircled in Figs.\ref{f2}(a),(b) by the red line, where stable patterns $P_{1,2}$ are easily excitable by the cw pump. The simultaneous presence of MI and alternating drift instability bands for our specific choice of the pump amplitude $h_0=1$ is essential for the dynamical transformations between $P_{odd}$ and $P_{even}$ families, and could not be anticipated \textit{a priori}. Indeed, for $h_0>1$ the MI region expands, but the stability domains shrink, and vice-versa for $h_0<1$ (not shown).
%These unique circumstances determine our choice of pump amplitude $h_0=1$.

%%%%%%%%%%%%%%%%%%%%%%%%
% TRANSITIONS % FIG. 3 %
%%%%%%%%%%%%%%%%%%%%%%%%
%
%\textit{Robust comb replication and erasure.---}
%
A sequence of stepwise pattern transitions of the type $P_N\rightarrow P_{N+1}$ ($P_{N-1}$) leading to deterministic self-replication (-erasure) are shown in Fig. \ref{f3}. Simulation of Eq.\ref{eq1} was initiated with $P_2$ pattern obtained at $\delta=0.84$, $\sigma_z=10$, $\sigma_x=2$, $h_0\approx9$. Here we take into account that in the experiment the pump profile is typically localized in $x$, hence pump amplitude, $h_0$, was adjusted to closely correspond to the dynamics observed for uniform in $x$ pump.
As time goes on, the pump width, $\sigma_z$, is gradually increased up to $60$, where excitation of stable $P_8$ pattern is observed, and then decreased back down to $10$, leading to restoration of stable $P_2$ pattern [see Figs. \ref{f3}(a) and \ref{f3}(b)] {(Note that the total pump power increases with $\sigma_z$, as $h_0$ is constant).} Transitions are triggered at the times [marked by red and blue dots] where varying $\sigma_z$ drives $P_N$ outside its stability region and into the region where it becomes unstable and starts to drift spontaneously upwards or downwards in $z$ [see gray (purple) shaded regions for $P_{even}$ ($P_{odd}$) families]. When $\sigma_z$ increases (decreases), the drift induces the transition $P_N\rightarrow P_{N+1}$ $(P_{N-1})$, as expected from Figs.\ref{f2}(a),(b). While axially \textit{locked} patterns have zero average axial position [c.f. Figs. \ref{f1} $(i)-(vii)$], $\langle z\rangle\equiv\int_0^{x_p}dx\int_{\infty}^\infty zdz|\psi-\psi_0|^2/U\equiv0$, the $z$-drifting transient states do not, and thus the transitions $P_N\rightarrow P_{N\pm1}$ are characterised by pronounced peaks in $\langle z\rangle$ vs $t$, apparent in Fig. \ref{f3}(b). These peaks show that transitions at larger $\sigma_z$ (larger $N$) take more time, in agreement with the fast decrease of the growth rate $\text{Re}(\lambda)$ with pump width [Fig.\ref{f2}(c)] (see, e.g., Ref.\cite{magg00} for discussion on drift speeds). Details of the transition $P_2\rightarrow P_3$ are illustrated within dashed rectangle in Fig.\ref{f3}(b) and selected transient pattern profiles are shown in Figs.\ref{f3} $(i)-(iii)$. We emphasize that {\it coexistence of a drift unstable family and a stable one is crucial\/} for the reported effects. Fortunately, stationary pattern states with nonzero average position, $\langle z\rangle\neq0$, were not found to exist within the investigated parameter space {(c.f. \cite{SM}, section V)}. States of this sort may exist with modulated background and, if existed, they could have frustrated comb replications.

%If a stable family did not exist, drift would not be arrested and transitions would result in a continuous axial scroll \cite{scr05apb}.%,scrCS1,scrCS2}.

%Results shown in Fig.\ref{f3} correspond to the parameters of a hollow silica glass cylinder of $R=100\ \mu$m and wall thickness $1.25\ \mu$m, pumped at $\lambda_p=1.55\ \mu$m, yielding $\tau=3.2$ ps, $B_{2,x}\approx1.8\times 10^{-6}$ and $B_{2,z}\approx1.37\times10^{-4}$ for the TE mode, and quality factor $Q\equiv\omega_p\tau/\gamma=10^6$ (realistic for cylinders \cite{sumcyl}). The closest to pump wavelength zero of the GVD is located below $\lambda=1\ \mu$m, thus $B_{2,x}$ dominates the dispersion.
% ($\omega^{(2)}\approx36\ \mu$m$^2$THz)
%For this geometry, spots are $\sim90\ \mu$m apart along $x$ and rows (combs) are spaced by $\sim0.5$ mm along $z$.
%($\Delta z=1$ corresponds to $\Delta Z\approx170\ \mu$m).%[c.f. Fig.\ref{f1},Fig.\ref{f3},Fig.\ref{f4}].

We note that the drift instability bands in Fig.~\ref{f3}(a) do not coincide exactly with those in Fig.~\ref{f2}(c), plotted also as an inset in Fig.~\ref{f3}(a), for clarity. This is because patterns in Fig.~\ref{f1} and Fig.~\ref{f2} were calculated for unit $\xi(x,t)$ [flat in $x$ pump], while the propagation takes into account the $x$-localization of $\xi(x,t)$. The impact of $\xi(x,t)$, representing a non-autonomous perturbation, is well tested in $1D$, but it is not in $2D$. Hence, our results show the robustness of the comb replication effect in the regime where the steady state calculations of Figs. \ref{f1} and \ref{f2} cannot be easily done, a feature that is crucial for the experimental demonstration of the phenomenon. We stress that other case-specific autonomous perturbations arising from linear dispersion \cite{erk,lam,milianOE,parra,milianPRL18,vahNatCom,braschSci,skryOE17,mbe17} and moderate Raman effect \cite{milianPRA,karpov16,vahalaStokes,gaetRam} are perfectly compatible with robust nonlinear states and, for the sake of generality, are not considered here.
 % modal index of pump is 1.36
%%%%%%%%%%%%%%%%%%%%%
%% FIG. 4. Spectra %%
%%%%%%%%%%%%%%%%%%%%%
%

%\textit{Spectra.---}
%
Figures \ref{f4}(a)-(c) show multi-comb spectra in the $(k_x,z)$ plane of patterns $P_1$ (a), $P_3$ (b), and $P_8$ (c), obtained under the same conditions, as those in Fig. \ref{f3} at constant $\sigma_z$ (see labels). One-dimensional spectra at specific $z$ values are shown in Figs.\ref{f4}(d)-(f). Because patterns include 7 periods in the $x$-direction [c.f. Figs.\ref{f3}(i)-(iii)], all spectra feature high-amplitude peaks separated by $7$ free spectral ranges (FSRs). The other much weaker resonances appear due to $x$-dependence of $\xi(x,t)$, as noted previously \cite{kartOE17}, and tend to zero in the case $\xi(x,t)=1$ \cite{SM}. In the geometry considered above, the spectra in Figs.~\ref{f4}(e,f) span from $1.35-1.82\ \mu$m, corresponding to an equivalent duration of individual solitons of about $\sim200$ fs.
%%%%%%%%%%%%%%%%%
%% Conclusions %%
%%%%%%%%%%%%%%%%%

%\textit{Concluding remarks.---}
%
In closing, we stress that the phenomenon described here reveals a mechanism that replicates and erases frequency combs along the axis of a cylindrical microresonator in a stepwise deterministic manner, affording a robust way to manipulate multi-frequency combs states in the form of spatiotemporal patterns. The frequency combs tend to be exact copies of each other and are inherently synchronised. These two features are promising for important applications such as spectroscopy, communications, RF links, and imaging. This effect is based on a fundamental drift instability that dynamically connects pattern families with odd and even number of rows (or combs), while preserving the overall symmetry. Emerging patterns via these transformations are locked and robust. Our results also bring fundamental understanding of the mechanism of pattern transformations, a phenomenon of major importance in the general context of nonlinear waves in dissipative media.
%
%%%%%%%%%%%%%%%%%%%%%
%% Acknowledgments %%
%%%%%%%%%%%%%%%%%%%%%
\begin{acknowledgments}
{\small This work was parially supported by the Government of Spain (grants IJCI-2016-27752, MTM2016-75963-P, FIS2015-71559-P; Severo Ochoa CEX2019-000910-S); Generalitat de Catalunya; CERCA; Fundaci\'{o} Cellex; and Fundaci\'{o} Mir-Puig.}
\end{acknowledgments}

%
% ADD HERE SUPP MAT

\clearpage
\counterwithin{figure}{section}

\section*{Supplemental Material: Reversible Self-Replication of Spatio-Temporal Kerr Cavity Patterns}
\appendix

\section{Theoretical model}

\subsection{Outline for derivation}

Our starting point is the wave equation for the complex electric field in a linearly dispersive and nonlinearly non-dispersive dielectric (non-magnetic) environment,
\begin{eqnarray}
&& \nonumber\ora\nabla\times\ora\nabla\times{\bf E}+\frac{1}{c^2}\partial_T^2(\epsilon_L({\bf r},T)\ast{\bf E})
=-\frac{1}{c^2}\partial_T^2{\bf P}^{\scaleto{nl}{5pt}}({\bf r},T,\bE)\\ &&=-\frac{\chi^{(3)}}{4c^2}\partial_T^2[2|\bE|^2\bE+\bE^2(\bE+\bE^*)] \label{eq:s1},
\end{eqnarray}
where $\ast$ denotes the convolution product and $\chi^{(3)}\equiv\hat\chi_{xxxx}^{(3)}=\frac{8}{3}n_0n_2$ $(\text{m}^2/\text{W})$, so intensity is $I\equiv|{\bf E}|^2$. The total permittivity is $\epsilon_L({\bf r})=\epsilon_c({\bf r})+\epsilon_e(\bar{\bf r})-\epsilon_{clad}$, where subscripts $c,e,clad$ denote cavity, external coupling device, and cladding. In addition, {$\epsilon_c=\epsilon_c^{re}+i\epsilon_c^l$, where $\epsilon_c^{re}$ and $\epsilon_c^l$ represent the real and imaginary (or lossy) parts of $\epsilon_c$, respectively}. The total field is decomposed into many axial and polar modes of the cylinder (we assume a single radial mode  only) and the one mode (or time-harmonic field) of the pump:
\begin{eqnarray}
&&
{\bf E}= \label{eq:s2}\\ &&\nonumber
\underbrace{\int_{-\infty}^\infty\frac{dk_z}{2\pi}\sum_{m=0}^\infty A_{m,k_z}(T){\bf\Upsilon}_{m,k_z}({\bf r})e^{-i\omega_{m,k_z} T}}_{\text{${\bf E}_c$, intracavity}}
+
\underbrace{p{\bf M}({\bar{\bf r}})e^{-i\omega_p T}
}_{\text{pump}},
\end{eqnarray}
${\bf\Upsilon}_{m,k_z}({\bf r})\equiv{\bf F}_{m,k_z}( r)e^{i(m\phi+k_zZ)}$ and ${\bf M}(\bar{{\bf r}})={\bf \Phi}(\bar{X},\bar{Y})e^{i\beta_p\bar{Z}}$ where ${\bf F}(r)$ and ${\bf\Phi}(\bar{X},\bar{Y})$ are the radial and transverse profiles of the cavity and driving fields, respectively. We use normalization $\int rdr|{\bf F}|^2\equiv1$. Modal indices are $m\equiv R n_{m,k_z}\omega_{m,k_z}/c\in\mathbb{N}$, representing the number of effective wavelengths ($\lambda_0/n_{m,k_z}$) within a microcavity roundtrip ($2\pi R$), and $k_z$, the axial propagation constant. Here, ${\bf r}\equiv\{r,Z,\phi\}$ are the natural cylindrical coordinates for the micro-cylinder and ${\bar{\bf r}}\equiv\{\bar{X},\bar{Y},\bar{Z}\}$ describe the driving field, with $\bar{X},\bar{Y}$ being transverse to the driving field propagation direction, $\bar{Z}$ ({note that $Z$ and $\bar{Z}$ represent different coordinates, see Fig. \ref{fig:S1}}). {$\beta_p$ is the propagation constant of the driving field at the laser wavelength.}

Substituting field in the form (\ref{eq:s2}) into Eq. (\ref{eq:s1}) and projecting onto a reference mode ${\bf\Upsilon}_{m',k_z'}({\bf r})$ leads to the rate equation for the amplitudes, $A_{m,k_z}(T)$. Finally, introduction of the field envelope
\begin{equation}
\nonumber
\Psi(Z,\phi,T)=\sum_{m=0}^\infty\int_{-\infty}^\infty\frac{dk_z}{2\pi}A_{m,k_z}(T)e^{-i\omega_{m,k_z}T+i(m-m_0)\phi+ik_zZ}
\end{equation}
in the equation for $\partial_TA_{m,k_z}$ yields the damped-driven nonlinear Schr{\"o}dinger equation in physical units,
%
%\begin{widetext}
\begin{eqnarray}
&&
\partial_T\Psi(T,\phi,Z) =-i\sum_{q=0}^\infty\frac{\omega^{(q,0)}}{q!}(-i\partial_\phi)^q\Psi+\label{eq:s3}
\\ &&\nonumber
+\frac{i}{2}\omega^{(0,2)}\partial_Z^2\Psi-\Gamma\Psi+i{\mathcal{V}}|\Psi|^2\Psi+ip{\mathcal{K}}(\phi,Z)e^{i\omega_pT}.
\end{eqnarray}
%\end{widetext}
%
$\omega^{(p,q)}\equiv\left.\partial_m^p\partial_{k_z}^q\omega(m,k_z)\right\vert_{(m_0,0)}$ are the dispersion coefficients and $\Gamma\left.\equiv\frac{\omega}{2}\langle{\bf \Upsilon}| \epsilon_c^l(\omega){\bf \Upsilon}\rangle\right|_{(m_0,0)}$ accounts for linear losses, related to the quality factor by $Q=\omega_p/\Gamma$. The nonlinear coefficient is defined as:
\begin{eqnarray}
&&
{\mathcal{V}}\equiv\frac{\chi^{(3)}\omega_p}{8\pi\epsilon_L}\left[\langle{\bf F}|(|{\bf F}|^2{\bf F})\rangle_{\{r\}}+\frac{1}{2}\langle{\bf F}|{\bf F}^2{\bf F}^{*}\rangle_{\{r\}}\right],\label{eq:V}
\end{eqnarray}
where bra-kets denote radial integrals within the nonlinear region. The pump distribution is given by
%\begin{widetext}
\begin{eqnarray}
&&
% EQNS FOR kappa's
{\mathcal{K}}\equiv
\frac{\omega_p}{2\pi}
\int_{-\infty}^\infty\frac{dk_z}{2\pi}\sum_{m=0}^\infty\mathcal{I}_V(m,k_z)e^{i(m-m_0)\phi+ik_zZ}
\label{eq:K}
,
\\ &&
\mathcal{I}_V(m,k_z)\equiv\langle{\bf  F}|\frac{\epsilon_c^{re}-\epsilon_{clad}}{2\epsilon_L}{\bf \Phi}e^{-im\phi-ik_zZ}e^{i\beta_p{\bar{Z}}}\rangle_V,
\label{eq:s5}
\end{eqnarray}
%\end{widetext}
where expression (\ref{eq:s5}) involves integral over the whole volume of the structure.

The above derivation procedure involved several standard approximations. {First, we make use of the paraxial approximation, which implies that only the first-order time derivatives are considered. Second, we assume a slow temporal variation of the pump field with respect to amplitudes $A_{m,k_z}(T)$. Third, the dispersion of the real modal refractive index is considered to be the main source for dispersion, which is a necessary assumption to derive the continuous model in Eq. \ref{eq:s3}. Fourth, the Kerr nonlinearity is the dominant one and all others are disregarded. Fifth, and last, we are considering the case when the pump excites predominantly the cavity mode with frequency $\omega_0\equiv\omega_{m_0,0}$; the closest resonance to the pump $\omega_p$ having zero axial group velocity, $k_z=0$}.

\subsection{Rewriting and normalizing Eq. (\ref{eq:s3})}

Next we employ the scaling that utilizes the physical cavity roundtrip time $\tau$ and cylinder's radius $R$:
\begin{eqnarray}
&&\nonumber t'=\frac{T}{\tau};\ z'=\frac{Z}{2\pi R}; x'=\frac{\phi-\omega^{(1,0)}\tau t'}{2\pi},\\ && \nonumber
\Psi=\frac{1}{\sqrt{{\mathcal{V}}\tau}}\psi'(t',z',x')e^{-i\omega_p\tau t'}
\end{eqnarray}
to cast Eq. (\ref{eq:s3}) in the following form:
%
%\begin{widetext}
\begin{eqnarray}
&&
\partial_{t'}\psi' =\nonumber -i\sum_{q=2}^\infty B_{q,x}(-i\partial_{x'})^q\psi'+iB_{2,z}\partial_{z'}^2\psi'+\\ &&
-(i\delta'+\gamma)\Psi'+i|\Psi'|^2\Psi'+ih'(x',z',t')
\label{eq:s8},
\end{eqnarray}
%\end{widetext}
%
with $\delta'=(\omega_p-\omega_0)\tau$, $\gamma=\Gamma\tau$, $h'=\tau\sqrt{\tau\mathcal{V}}p\mathcal{K}$, and
\begin{eqnarray}
&& B_{q,x}\equiv\frac{\tau}{(2\pi R)^qq!}\left.\frac{\partial^q\omega}{\partial{k_x}^q}\right|_{(k_x=m_0/R,k_z=0)},\\ &&  B_{2,z}\equiv\frac{\tau}{(2\pi R)^qq!}\left.\frac{\partial^2\omega}{\partial{k_z}^2}\right|_{(k_x=m_0/R,k_z=0)}.
\end{eqnarray}
Note from the above definition that $B_1\equiv1$ because the physical group velocity, $v_g^{ph}\equiv\partial_{k_x}\omega\vert_{(m_0/R,0)}$ is exactly $2\pi R/\tau$. Dispersion coefficients $B_{2,x}$, $B_{2,z}$ can be eliminated by rescaling of coordinates $x'\rightarrow x'\sqrt{2B_{2,x}}$ and $z'$ $\rightarrow$ $z'\sqrt{2B_{2,z}}$. Also, we note that Eq. (\ref{eq:s8}) is invariant to the global scaling $\{t,x,z,\psi,b_{N,x},\bar{\gamma},\delta,h\}=$ $\{t'a,x'\sqrt{a},$ $z'\sqrt{a},\psi'/\sqrt{a},$ $B_{N,x}a^{N/2-1},\gamma/a,\delta'/a,h'a^{-3/2}\}$ {($N\in\mathbb{N}\geq1$)}. %(note that $b_{2,x}=B_{2,x}$).
Because our system is always dissipative in practice, the scaling $a=\gamma$ does not introduce singularities and transforms Eq. (\ref{eq:s8}) into its final dimensionless form
%
%
%\begin{widetext}
\begin{eqnarray}
&&\nonumber
\partial_{t}\psi = -i\sum_{q=2}^\infty \frac{b_{q,x}}{(2B_{2,x})^{q/2}}(-i\partial_{x})^q\psi+\frac{i}{2}\partial_{z}^2\psi+\\ &&
-(i\delta+1)\Psi+i|\Psi|^2\Psi+ih(x,z,t)
\label{eq:s9},
\end{eqnarray}
%\end{widetext}
%
which is identical to Eq. (1) from the main text with the addition of the higher dispersion terms (those with $q\geq3$) accounting for the full cylinder's dispersion along $x$. In main text we restricted our discussion to dispersion terms up to second order, because fixing the higher order dispersion (HOD) in a meaningful way implies selecting a particular cavity geometry (see also Appendix B), while our aim is to keep the discussion as general as possible. In addition, conversely to single pass systems, HOD (as well as other higher-order nonlinear effects) do not compromise the existence and robustness of the stationary nonlinear states, as discussed in main text. 
\subsection{A remark on normalization and MI}

Physically, one motivation to work with the normalized Eq. (\ref{eq:s9}) is that when the background field is flat, $h(x,z)=$const, the pump value $h=1$ excites a nonlinear resonance with peak intensity $\text{max}\{|\psi|\}=1$, which, in turn, coincides with the modulational instability (MI) threshold for the background with anomalous GVD ($B_{2,x}>0$). Thus, $h>1$ means that a flat background with MI always exists (as the field modulus takes the values $|\psi|>1$ for some finite interval in $\delta$), while for $h<1$ there is no MI at all (field modulus $|\psi|<1$ for all $\delta$).

The above picture changes slightly when $h$ is a function of $z$, as in our main text. Indeed, Figs. 2(a) and 2(b) in the main text exhibit a finite MI region around $\sigma_z\in[2,15]$ despite the pump amplitude $h_0=1$. The reason why the MI threshold depends upon $\sigma_z$ at a fixed $h_0$ is because the diffraction term, $\sim\partial_z^2\psi$, must be kept when computing background states, whose amplitude then naturally depends on $\sigma_z$. Note that we have chosen $h_0=1$ for Figs. 1 and 2 in the main text, that in the flat pump case ($h=$const) corresponds to the exact threshold for MI. This is why MI disappears in Figs. 2(a) and (b) when $\sigma_z\rightarrow\infty$. One could have expected the MI region to close asymptotically as $\sigma_z \rightarrow \infty$, but in reality the system is such that MI disappears at finite $\sigma_z$ values (around $15$).

\subsubsection{Scaling summary}
The relation between variables introduced in Eqs. (\ref{eq:s9}), (\ref{eq:s8}), and (\ref{eq:s3}), respectively, can be summarized in the following scaling summary:

\begin{widetext}
\begin{eqnarray}
&& \nonumber
t=t'\gamma=\frac{T\gamma}{\tau},\quad\quad\quad\quad\quad\quad x=x'\sqrt{\frac{\gamma}{2B_{2,x}}}=\frac{X}{2\pi R}\sqrt{\frac{\gamma}{2B_{2,x}}},\quad\quad\quad\quad\quad\quad z=z'\sqrt{\frac{\gamma}{2B_{2,z}}}=\frac{Z}{2\pi R}\sqrt{\frac{\gamma}{2B_{2,z}}},\quad\\ &&
b_{N,x}=B_{N,x}\gamma^{N/2-1}=\left.\frac{\partial^N\omega(k_x,k_z)}{\partial k_x^N}\right|_{\frac{m_0}{R},0}\frac{\tau\gamma^{N/2-1}}{(2\pi R)^NN!},\quad\quad\quad\quad\quad\quad\quad\quad\quad \delta=\frac{\delta'}{\gamma}=\frac{(\omega_p-\omega_0)\tau}{\gamma},\label{eq:scaling}
\\ && \nonumber
\psi=\frac{\psi'}{\sqrt{\gamma}}=\sqrt{\frac{\mathcal{V}\tau}{\gamma}}\Psi e^{i\omega_pT}\approx\sqrt{\frac{\mathcal{V}\tau}{\gamma}}\langle  {\bf F}(r)|{\bf E}_c(r,Z,\phi,t)\rangle_{\{r\}}e^{i\omega_pT-im_0\phi},\quad %\\ &&
\quad h=\frac{h'}{\gamma^{3/2}}=\left(\frac{\tau}{\gamma}\right)^{3/2}\mathcal{V}^{1/2}p\mathcal{K},
\end{eqnarray}
\end{widetext}
with $\gamma=\omega_p\tau/Q$, $\langle{\bf F}|{\bf F}\rangle_{\{r\}}\equiv1$. ${\bf E}_c$, $\mathcal{V}$, and $\mathcal{K}$ are given in Eqs. (\ref{eq:s2}), (\ref{eq:V}), and (\ref{eq:K}), respectively.

\section{Physical estimates}

In this section we provide realistic estimates for parameters of cylindrical microresonators and driving laser beams.

\begin{figure}
    \centering
    \includegraphics[width=.5\textwidth]{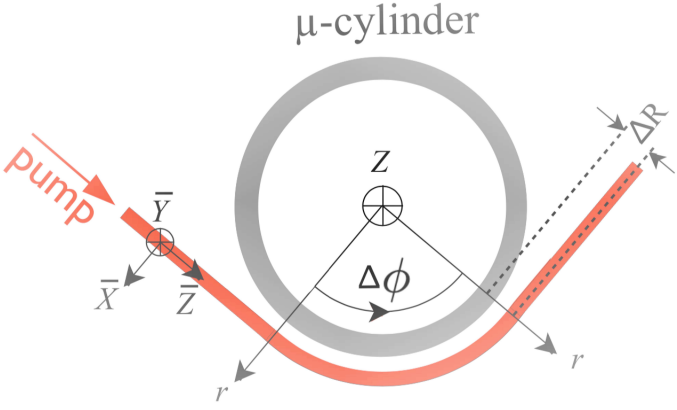}
    \caption{Top view of the cylindrical microresonator and pump waveguide feeding the cavity over the arc $\Delta\phi$. {The $\bar{Z}$ axis corresponds to the path of the bus waveguide (in red), while $\bar{X}$ and $\bar{Y}$ are perpendicular to it.}}
    \label{fig:S1}
\end{figure}

In order to present estimates, we need to refer to a particular pumping geometry, i.e., we need to specify the actual trajectory of the pump, $\bar{Z}$, in relation to the ${r,Z,\phi}$ coordinates describing the cylindrical microresonator. In what follows, we consider the situation sketched in Fig. \ref{fig:S1}, where the pump field propagates in the ${r,\phi}$ plane over a circular arc of radius $R+\Delta R$. $\Delta R$ is the radial distance between the center of the waveguides in the coupling region.
\begin{figure}
    \centering
    \includegraphics[width=.5\textwidth]{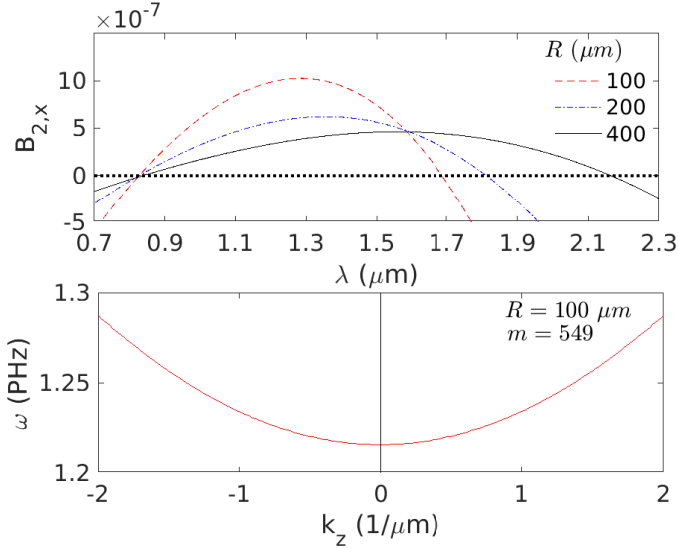}
    \caption{(a) GVD for the three micro-cylinders characterised in Table\ref{tab:1}. (b) Parabolic dispersion around the cylinder's cut-off ($k_z=0$) for the resonance $m=549$ of the $R=100\ \mu$m cylinder, corresponding to $\lambda_0=1.55\ \mu$m. Profiles for the other two cylinders are almost identical.}
    \label{fig:S2}
\end{figure}
%\begin{widetext}
\begin{table*}[]
\begin{tabular}{|c|c|c|c|c|c|c|c|c|c|c|}
\hline
\begin{tabular}[c]{@{}c@{}}$R$\\ ($\mu$m)\end{tabular} & \begin{tabular}[c]{@{}c@{}}$w$\\ ($\mu$m)\end{tabular} & \begin{tabular}[c]{@{}c@{}}zero GVD wavelengths:\\ {[}$\lambda_1,\lambda_2${]}\\ ($\mu$m)\end{tabular} & $m_0$  & \begin{tabular}[c]{@{}c@{}}$\tau$\\ (ps)\end{tabular} & \begin{tabular}[c]{@{}c@{}}$Q$\\ ($\times10^6$)\end{tabular} & \begin{tabular}[c]{@{}c@{}}pulse width, $\Delta t$\\ (fs)\end{tabular} & \begin{tabular}[c]{@{}c@{}}$P/z$\\ (W/mm)\\ $h_0=1$,\\ $\Delta\phi=2\pi$\end{tabular} & \begin{tabular}[c]{@{}c@{}}$P/z$\\ (W/mm)\\ $h_0=9$, \\ $\Delta\phi=\pi/8$\end{tabular} & $B_{2,z}$ & \begin{tabular}[c]{@{}c@{}}$\Delta z=1$\\ ($\mu$m)\end{tabular} \\ \hline
100    & 1.07   & {[}0.8,1.7{]}  & 549  & 3.2   & 3.8  & $\sim100$  & 0.25  & 19       & $1.5\times10^{-4}$ & 344     \\ \hline
200    & 1.12   & {[}0.8,1.8{]}  & 1102 & 6.4   & 7.6  & $\sim200$  & 0.046 & 3     & $7.5\times10^{-5}$ & 490     \\ \hline
400    & 1.25   & {[}0.8,2.2{]}  & 2224 & 12.7  & 15   & $\sim400$  & 0.03 & 2     & $3.7\times10^{-5}$ & 690     \\ \hline
\end{tabular}
\caption{Realistic parameters associated to geometries where the reported spatiotemporal patterns could be observable. These parameters are obtained with a dispersive Maxwell solver in cylindrical coordinates. Pump wavelength is fixed at $\lambda_p=1.55\ \mu$m, GVD of the cylinder at $B_{2,x}=5\times10^{-7}$, and the material considered is silica glass. The bus waveguide is assumed to have the same width, $w$, as the cylinder.\label{tab:1}}
\end{table*}
%\end{widetext}
%
Such choice of the pump geometry implies that over the coupling region, the bus waveguide has the trajectory given by $\bar{Z}=\phi(R+\Delta R)$ and the mode of the bus waveguide lies on the $\{r,Z\}$ plane, so $\bf\Phi$($\bar{X},\bar{Y})\rightarrow{\bf\Phi}(r,Z)$. In this case, Eq. (\ref{eq:s5}) becomes:
\begin{eqnarray}
 &&
\mathcal{I}_V(m,k_z)=\langle{\bf  F}|\frac{\epsilon_c^{re}-\epsilon_{clad}}{2\epsilon_L}{\bf \Phi}e^{-ik_zZ}\rangle_{\{r,Z\}}\mathcal{I}_{m},\label{eq:IV}
\\ &&\nonumber
\mathcal{I}_{m}\equiv\int_0^{\Delta\phi}d\phi e^{-i(m-\beta_p[R+\Delta R])\phi}=\\ &&
=\left\{
\begin{array}{cc}
i\frac{[e^{-i(m-\beta_p[R+\Delta R])\Delta\phi}-1]}{m-\beta_p[R+\Delta R]} &,\ m\neq\beta_p[R+\Delta R]  \\
\Delta\phi &,\ m=\beta_p[R+\Delta R]
\end{array}
\right.\label{eq:Im}
.
%\label{eq:s5}
\end{eqnarray}
For the power estimates provided below (see Table \ref{tab:1}), we assume the simple case where the pump mode is uniform along $Z$, $\Phi(\textbf{r},Z)\rightarrow \Phi(\textbf{r})$. This approximation will lead to power levels which will differ from the actual values by standard form factors, depending on the particular shape of the pump beam (Gaussian, super-Gaussian or other higher-order modes of the bus waveguide). Thus results below are representative of real situations and should be taken as order of magnitude estimates.

We proceed by fixing the pump wavelength to $\lambda_p=2\pi c/\omega_p=1.55\ \mu$m and a reference loss parameter to $\gamma\equiv\omega_p\tau/Q=0.001$. Then, the span $\Delta x=32$ in our simulations (see Figs. 1 and 3 in the main text) corresponds to $\Delta X=2\pi R$, and thus $32=\sqrt{\gamma/[2B_{2,x}]}$ [see scaling for $x$ in Eq.(\ref{eq:scaling})] that fixes the GVD parameter to: $B_{2,x}\approx5\times10^{-7}$. With the aid of a Maxwell solver, we find realistic geometries (hollow cylinders made of silica glass with radius $R$ and wall width $w$) in which such GVD is found [see Fig. \ref{fig:S2}(a)]. Results for this geometry are shown in table\ref{tab:1}. The process to build the table is as follows:

\begin{itemize}
    \item Running Maxwell solver in cylindrical coordinates with fixed $R$, $w$ and $\lambda_0=\lambda_p=1.55\ \mu$m, we get the profile $\bf F(r)$ (and normalise it to $\langle {\bf F}|{\bf F}\rangle_{\{r\}}\equiv1$), $m_0$, $\tau=2\pi R/v_g$, and $Q$.
    \item By scanning in $\lambda_0$ we obtain the dispersion, $B_{2,x}(\lambda_0)$ [see Fig. \ref{fig:S2}(a)]. Temporal duration of solitonic pulses in the patterns can be estimated using $\Delta t\approx \tau \Delta X/(2\pi R)=\tau\sqrt{2B_{2,x}/\gamma}\Delta x$, where the full width at half maximum $\Delta x$ can be measured directly from our numerical data, see, e.g., Figs. 1 and 3 in the main text.
    \item $\bf \Phi$ and $\beta_p$ are computed for another cylinder or radius $R+\Delta R$ and same wall width $w$. Note from Fig. \ref{fig:S1} that this is a reasonable assumption.
    \item $\bf F$, $\bf \Phi$, $\beta_p$, $R+\Delta R$ and the chosen $\Delta\phi$ are used to compute $\mathcal{I}_V$ in Eq. (\ref{eq:IV}) for a fixed $m$ and $k_z=0$, which is inserted in Eq. (\ref{eq:K}) to obtain the pump distribution within the cylinder, $\mathcal{K}$. Span of values of $m$ used in the sum in Eq. (\ref{eq:K}) is large enough to ensure convergence.
    \item The pump intensity $|p|^2$ is computed from $p=h_0\gamma^{3/2}/[\tau^{3/2}\mathcal{V}^{1/2}\text{max}\{|\mathcal{K}|\}]$.
    \item The pump power per unit length along $z$ is then obtained as $P/z\equiv \mathrm{w}_r|p|^2$, where $\mathrm{w}_r\equiv\int_0^\infty dr|{\bf F}|^2/\mathrm{max}\{|{\bf F}|^2\}$ is the modal width along $r$.
    \item Finally, executing the Maxwell solver with fixed $m$ and scanning over $k_z$ gives $\omega(m_0,k_z)$ from which $B_{2,z}$ is computed [see Fig. \ref{fig:S2}(b)]. The latter provides the conversion factor allowing to transform dimensionless width of solution along the $z$-axis into physical units.
\end{itemize}

As an example, the patterns $P_7$ require $\sigma_z\approx50$ for their existence and stability [see Fig. 2(a) in the main text]. For the $R=200\ \mu$m cylinder, this corresponds to a physical $z$-width of $50\times490\ \mu m\approx2.5$ cm. In the localised pump case, $\Delta\phi=\pi/8$, we will then need a total beam power of $3$ W/mm $\times 25$ mm $=75$ W. Such power levels, even if not optimised here, are attainable with CW fiber lasers.

From the above parameter list, it is clear that the increase of the cylinder's radius is beneficial as it decreases threshold pump powers, but it is detrimental for the temporal pulse widths and for the vertical separation between pattern rows. As noted, increasing $R$ also shifts the zeros of the GVD away from the pump wavelength, see Fig. \ref{fig:S2}(a), and thus HOD becomes more and more negligible. Because particular values of the HOD may be used only when studying a particular geometry, we did not include them in any of our simulations shown in the main text. We note, however, that addition of HOD is not a problem for the existence or robustness of solitonic frequency combs, as is well-known.

Regarding dispersion along the $z$-axis, we note that it is strongly parabolic, as reflected in Fig. \ref{fig:S2}(b), around modes of interest with $k_z=0$, as they orbit around the cylinder's circumference. The parabolic shape implies that the first-order derivative $\partial_{k_z}\omega(k_z=0)\equiv0$, and that the coefficient $\omega^{(0,2)}\equiv\partial^2_{k_z}\omega(k_z=0)>0$. This dependence also shows that higher-order dispersion terms in $z$ can be safely neglected. These facts make it obvious that the dispersion term considered in our model, $\propto\omega^{(0,2)}\partial^2_Z\Psi$ is adequate and not a mere idealisation.

Further optimisation of the realistic geometries, i.e., additional tuning of $R$, $Q$, pulse duration and threshold powers, can be easily performed by changing the span $\Delta x$ used in the main text by an integer number of pattern periods $x_p=32/7$, i.e., by considering patterns with larger or fewer number of spots along the $x$-axis. Extrapolation to different spans $\Delta x$ is meaningful provided that patterns with different number of spots behave dynamically in very similar ways. Our stability analysis suggests that this is indeed the case for patterns with $n>3$ periods along $x$.

Finally, we note, that for a given $R$, there exist other values of wall thickness, $w\in[4,5.3]\ \mu$m, for which the similar $B_{2,x}$ can be obtained. These values, however, are not included here, as they required unrealistically large pump powers due to the very small overlap, $\mathcal{I}_V$, between cavity and bus modes.

\section{Pump modulation along x}

\subsection{About the function $\xi(x,t)$}

The function $\xi(x,t)$ was introduced in the main text to account for the realistic non-uniform in $x$ pumping conditions, in which patterns in the frame travelling with group velocity see localized pump as a \textit{pulse} passing through them once per roundtrip. This pulse was assumed to be of a Gaussian shape, but it can be of any suitable form. From the scaling above, it is clear that the function (as introduced in main text)
\begin{equation}
\xi(x,t)\equiv\sum_{m=-\infty}
^\infty\exp{(-[x-m\Delta x-v_gt]^2/\sigma_x^2)}
\end{equation}
with $x\in[-\Delta x/2,+\Delta x/2)$, may be written in physical units as:
\begin{equation}
\xi(X,T)\equiv\sum_{m=-\infty}
^\infty\exp{(-[X-m2\pi R-v_g^{ph}T]^2/\sigma_X^2)},
\end{equation}
with $X\in[-\pi R,+\pi R)$, where the physical group velocity, $v_g^{ph}\equiv \partial_{k_x}\omega(m_0/R,0)$, is related to the normalised one in main text, $v_g$, by:
\begin{equation}
v_g^{ph}=v_g\frac{2\pi R}{\tau}\sqrt{\frac{2B_{2,x}}{\gamma}}.
\end{equation}
Mathematically, the function $\xi$ represents a periodic train of pulses spaced by $\Delta X=2\pi R$. It can be easily seen that only one of these pulses is in the cavity, i.e., in the interval $X\in[-\pi R,+\pi R)$ and that each pulse goes through the cavity just once. Hence, physically, the function $\xi$ correctly describes the action of spatially localized pump.

\subsection{Effect of localized pump on multi-comb spectra}

\begin{figure}
    \centering
    \includegraphics[width=.5\textwidth]{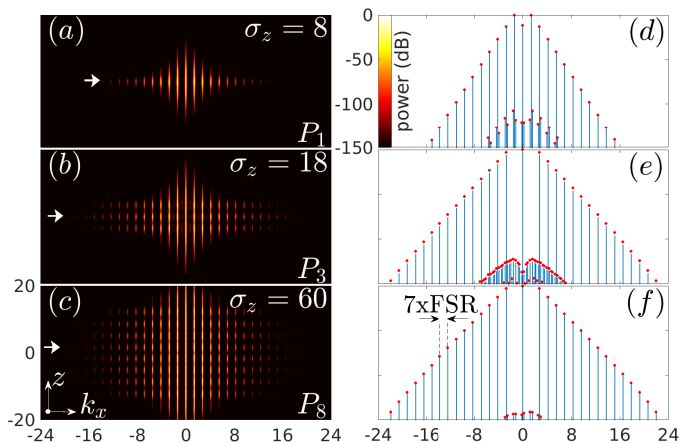}
    \caption{Spectra of multi-frequency comb states with a flat in $x$ background field. This figure is analogous to Fig. 4 in main text: Spectra along the cylinder's axis, $z$, for (a) $P_1$, (b) $P_3$, and (c) $P_8$ ($\sigma_z$ values are indicated in each plot) for flat pump in $x$: $h_0=1$ ($\xi=1$). (d,e,f), spectral slices at the selected $z$-positions (white arrows) in (a,b,c), respectively. Spectral power is shown over the range $[-150,0]$ dB in all plots. The background field, $\psi_0(x,y)$, was subtracted prior to Fourier transforms. Correspondence to physical units: $\Delta\kappa_x=1\Leftrightarrow 1/40$ $\mu \text{m}^{-1}$. At $\lambda_p=1.55$ $\mu$m, $\Delta\kappa_x=48\Leftrightarrow\lambda_0\in[1.35,1.82]\mu$m. $\Delta z=40\Leftrightarrow1.96$ cm.
    \label{fig:S3}}
\end{figure}

Fig. 4 of the main text showed spectra of selected spatiotemporal Kerr cavity patterns, corresponding to multi-frequency comb states. In that case, the pump field was considered localised along $x$. Here, we show in Fig. \ref{fig:S3} the equivalent spectra, but with flat background along $x$. Direct comparison between the two reveals that pump localisation along $x$, which introduces a global envelope along $x$, does impact the amplitude of some comb lines. In Fig. \ref{fig:S3} we observe some comb lines around $k_x=0$, $k_x\in{-8,8}$, at the level of $\lesssim100$ dB. These lines should actually be zero for uniform in $x$ pump, and they depart from zero only due to numerical noise. When the pump is modulated along $x$ these lines grow substantially. As seen in Fig. 4 of the main text, these lines generally remain between $5-50$ dB below the main comb lines, which are separated by 7 FSRs, since corresponding patterns include 7 periods along $x$. Characteristic features and extent of two types of spectra, with modulated and uniform in $x$ pumps, are nevertheless very similar.

\section{Breathing Dynamics}

\begin{figure}
    \centering
    \includegraphics[width=.5\textwidth]{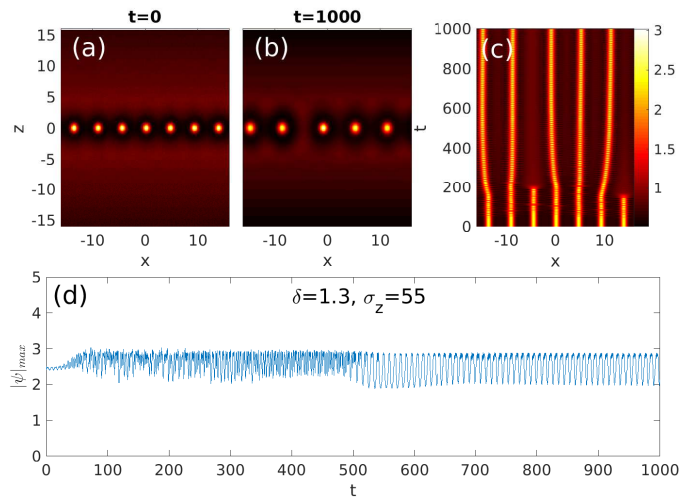}
    \caption{Propagation of a Hopf-unstable $P_1$ pattern. (a) input state with 7 periods, (b) output state with 5 periods, (c) temporal evolution of the field at $z=0$ vs time. (d) Maximum of the amplitude along propagation.}
    \label{fig:S4}
\end{figure}

In the main text we focus our attention on the interval in detuning, $\delta$, where drift was the only existing type of instability, so that drift occurred neatly and allowed us to replicate and erase frequency combs. However, the system, as expected, is far richer than that. In particular, we show here in Fig. \ref{fig:S4} the role of the Hopf instability, which is the dominant instability for $\delta\gtrsim 0.9$. As an example, we consider here the dynamics of the single-row pattern, $P_1$, with 7 periods at $\sigma_z=55$ and $\delta= 1.3$ [Fig.\ref{fig:S4}(a)]. This pattern has both Hopf and drift instabilities with growth rates $\sim0.08$ and $\sim0.01$, respectively. One could expect that the Hopf instability first induces breathing and then drift causes pattern motion along $z$. However, the strong perturbations induced by Hopf instability along propagation [Fig. \ref{fig:S4}(c)] transform the unstable $P_1$ patterns into another, different $P_1$ pattern with 5 periods [Fig. \ref{fig:S4}(b)]. The number of pattern periods at the output is stochastic and depends on the particular run. Once this transition into a new family happens, subsequent dynamics depends upon their stability properties. We did not attempt in this work to unveil all these details as they belong to different dynamical regimes than the one of our main focus. Such studies are thus left for future works.

\section{Norm of patterns versus pump width}
\begin{figure}
    \centering
    \includegraphics[width=.5\textwidth]{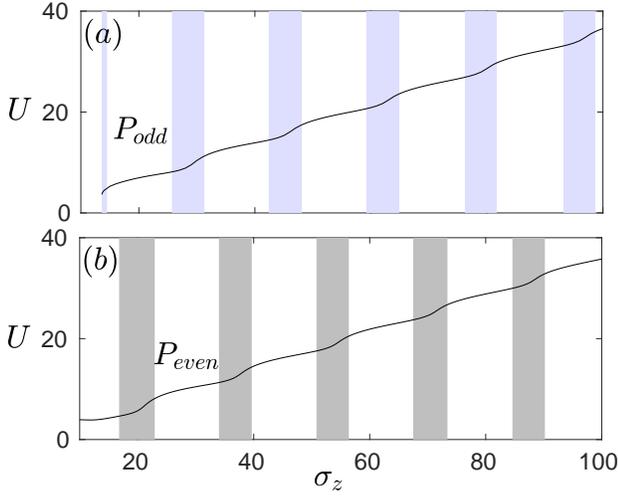}
    \caption{{Norm, $U$, vs pump width, $\sigma_z$, of the patterns at $\delta=0.84$ for the families with an (a) odd, $P_{odd}$, and (b) even, $P_{even}$, number of rows. Shaded areas represent the drift unstable regions, marked also in Fig.2(c) in the main text.}}
    \label{fig:S5}
\end{figure}
{
Figure \ref{fig:S5} shows the pattern branches in the norm vs pump width plane at $\delta=0.84$, the value of $\delta$ corresponding to Figs. 2(c) and Fig. 3 in the main text. The regions marked by the shaded areas in Figs. \ref{fig:S5}(a) and (b) correspond to the drift unstable regions for $P_{odd}$ and $P_{even}$ families, respectively, as discussed in main text. These branches illustrate explicitly the fact that, at a fixed $\sigma_z$, a drift unstable states of the $P_{odd}$ family can only meet stability by reshaping themselves into the coexisting $P_{even}$ states, and vice-versa, and hence the expected pattern transitions are of the type $P_{N}\rightarrow P_{N\pm1}$, depending on whether $\sigma_z$ is increased or decreased (c.f. Fig. 3 in main text). This picture, as discussed in the main text, is due to the \textit{alternated} stability domains of the two families.
}

{
In addition, the absence of saddle node bifurcations in Figs.\ref{fig:S5}(a) and \ref{fig:S5}(b) makes it clear that there is no coexistence of stable and unstable $P_{odd}$ states, nor coexistence of stable and unstable $P_{even}$ states. Thus transitions within the same pattern family, e.g., transitions of the type $P_{N}\rightarrow P_{N\pm2}$, cannot occur at a fixed $\sigma_z$. Such transitions could only take place if the rate at which $\sigma_z$ is varied is much higher than the rate at which drift takes place.
}

{
We emphasize that the transient states that appear dynamically during the transitions $P_{N}\rightarrow P_{N\pm1}$ (c.f. Fig. 3(i)-(iii) in main text) are asymmetric (in $z$) states with a nonzero velocity along $z$. While such states may be found explicitly as stationary solutions (with drift) in the case of uniform pump $\sigma_z\rightarrow\infty$ (e.g., Rung states), they are forbidden in our case because the modulated pump (along $z$) frustrates their existence as \textit{stationary} solutions. Thus transient states are essentially \textit{dynamical} states and it is not clear at this point how to associate them to branches of the forms $U$ vs $\delta$ or $U$ vs $\sigma_z$.
}

{
The picture described here holds for detunings in the range around $\delta\in[0.75,0.9]$ and $\sigma_z\gtrsim10$ (c.f. Fig. 2 in main text). For $\delta\gtrsim0.9$, branches start to exhibit snaking (c.f. Fig. 1(c) in main text), and thus branches analogous to those in Fig. \ref{fig:S5} do exhibit saddle nodes (not shown). In that case, other transitions may well occur, as those of the type $P_{N}\rightarrow P_{N\pm2}$. In that region of the parameter space, however, instabilities such as Hopf (see Appendix D above) make the picture more complex as patterns do also transit into other families with different pitch along $x$.
}
\end{document}